\pdfoutput=1
 
\documentclass[conference,fleqn]{IEEEtran}
\usepackage[cmex10]{amsmath}
\usepackage{amssymb}
\usepackage{amsthm}
\usepackage{mathtools}
\usepackage[pdftex]{graphicx}
\usepackage[font=footnotesize]{subfig}
\usepackage[linesnumbered,ruled,vlined]{algorithm2e}
\usepackage[usenames,dvipsnames,svgnames,table]{xcolor}
\usepackage{cite}
\usepackage{flushend}
\usepackage{booktabs}
\usepackage{mdwlist}

\usepackage{hyperref}
\hypersetup{%
  pdftitle={SGPD Volume Maximization for Community Detection},
  pdfauthor={Kasra Manshaei, Christian Bauckhage},
  pdfsubject={network science, graph theory},
  pdfkeywords={Peripheral Vertices, Community Detection, Graph Clustering, Graph Theory, Complex Networks},
  colorlinks=true,
  urlcolor=blue,
  citecolor=green,
  bookmarks=false}

\newtheorem{obs}{Obervation}

\newcommand{\argmax}{\arg\!\max}

\newtheorem{mydef}{Definition}
\graphicspath{{./}}

\begin{document}

\title{SGPD Volume Maximization for Community Detection}

\author{%
\IEEEauthorblockN{Kasra Manshaei}
\IEEEauthorblockA{University of Bonn \\ 53117 Bonn, Germany}
\and
\IEEEauthorblockN{Christian Bauckhage}
\IEEEauthorblockA{Fraunhofer IAIS \\ 53754 St.Augustin, Germany}
}

\maketitle

\begin{abstract}
In this note we briefly study the feasibility of community detection in complex networks using peripheral vertices. Our method suggests a novel direction in axiomizing the problem of clustering in graphs and complex networks by looking at the topological role each vertex plays in the community structure, regardless of the attributes. The promising strength of pseudo-peripheral vertices as a lever for analysis of complex networks is also demonstrated on real-world data.
\end{abstract}

\section{Introduction}

The explosion of data in 21\textsuperscript{st} century drew more attention to the field of data analysis and brought many questions on. Data storage devices are being improved dramatically to store massive offline data and high-speed internet connections have resulted in flow of incredible amount of online data. One of the most important parts of this digital data can be found in the form of \emph{relations}.
 Analysis of relational data is interesting in different research fields such as Computational Biology \cite{10.1371/journal.pcbi.1001131, bio} (investigating the community structure of biological networks), Sociology \cite{Christakis22072014, Borgatti13022009} (studying the interaction between individuals in a society), Physics of Complex Networks \cite{PhysRevE.80.056117, PhysRevE.74.016110} (analysis of statistical and structural properties of complex systems) and Computer Science \cite{Chakrabarti:2006:GML:1132952.1132954} (social network analysis). 

From a mathematical point of view relational structures are even older as Leonard Euler started the Graph Theory in 18\textsuperscript{st} century and since then this field has increasingly attracted attention. In this note, we'll look at one of the rarely covered points of view in graph theory, that is the feasibility of graph and network analysis using \textit{Peripheral Vertices}. This section continues with the definition of peripheral and pseudo-peripheral vertices and some background in this regard. The proposed approach to the problem of Community Detection will be discussed in Section \ref{sec:comm}. Section \ref{sec:expcomm} describes some experimental results on Wikipedia graph and finally, conclusions and some ideas for the future works are discussed in Section \ref{sec:conc}.

\subsection{Previous works}
Peripheral vertices in graphs have been the subject of some researches, mostly in Math community, however it has not been a mainstream research direction in the field. \cite{Chartrand90}, \cite{DBLP:journals/dm/ChartrandEJZ03}, \cite{Reid:1992:PEV:2712475.2712956}, \cite{Kyš2000} and \cite{bielak1983peripheral} focused on peripheral vertices from a pure graph theoretic point of view, \cite{ceaceres2003convex} and \cite{balakrishnan2008convex} looked at the peripheral vertices (also called \textit{antimedian} vertices in the latter) to study the convex subgraphs and convex sets of vertices in a graph. \cite{Gu93} studies the properties of graph $H$ which is constructed by embedding of two graphs $G_{1}$ and $G_{2}$ such that $G_{1}$ becomes a subgraph of $H$ containing all its peripheral vertices and $G_{2}$ is another subgraph in which the eccentricity of all vertices are equal to the radius of $H$. 

All references mentioned above are indeed contributions to the pure mathematics rather than anything else. The closest method to our research is probably the \textit{Archetypal Analysis} \cite{cutler} which is not designed to study graphs and complex networks but standard data. Archetypal Analysis, to the best of our knowledge, is the first algorithm in statistical data analysis which considers the furthest points in a dataset as levers of analysis. Archetypal Analysis proposes a matrix factorization algorithm based on the data points lying on the convex hull of the dataset which are the closest concepts to peripheral vertices of a graph. \cite{Bauckhage2015-AAA} and \cite{icpr14} propose an efficient version of Archetypal Analysis and a kernelized version of this algorithm respectively.
\subsection{On Pseudo-Peripheral Vertices}

Let $G=(V,E)$ be a simple connected graph with the vertex set $V$ and edge set $E$. For every vertex $v\in V$ the eccentricity $\epsilon(v)$ is defined as the maximum of its shortest geodesic distance to any other vertex in $G$. Peripheral vertices in a graph $G$ are the vertices with highest eccentricities. In other words, as the diameter of $G$ is the longest shortest path between any two vertices in $G$, two endpoints of a diameter are peripheral vertices. Another interpretation is to consider peripheral nodes as the rows and columns associated with the maximum entries in the distance matrix of $G$. These are actually the \textit{furthest} vertices in $G$. 
\begin{obs}
\label{obs:1}
Every connected simple graph has at least 2 peripheral vertices.
\end{obs}
\begin{obs}
\label{obs:2}
Any graph has a fixed number of peripheral vertices which is induced from its topology.
\end{obs}
In other words, the distance matrix of a connected simple graph has a fixed number of global maxima.

\begin{figure}[htb!]
  \centering
  \centerline{\includegraphics[width=8.5cm]{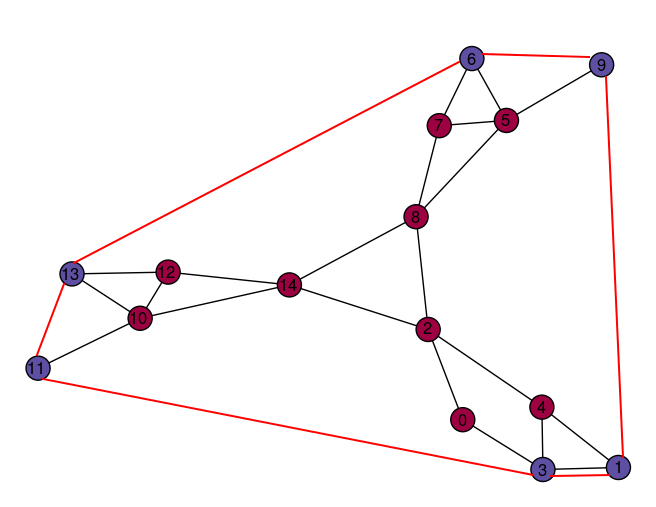}}
\caption{\label{fig:PerNodes}An illustration of peripheral nodes (blue ones) and the virtual  \textit{convex polygon} connecting them (red edges).}
\end{figure}

An intuitive interpretation of peripheral vertices is illustrated in Fig.~\ref{fig:PerNodes} where peripheral vertices are \textit{k} vertices in the graph maximizing the volume of a virtual convex polygon built by connecting those nodes. Readers may note that the convex polygon in Fig.~\ref{fig:PerNodes} is not really a geometric object as graphs are relational structures. This polygon is just to intuitively show that peripheral vertices should be \textit{as far as possible} so we can imagine them maximizing the volume of such a virtual shape.

The Observation \ref{obs:2} limits a peripheral node finder algorithm to a fixed number of peripheral vertices. To cope with this constraint, we continue with definition of \textit{pseudo-peripheral} vertices.
\begin{mydef}
\label{def:1}
A vertex $u\in V$ is called pseudo-peripheral if\\
\begin{center}
$|\epsilon(u)-diam(G)|<\delta$
\end{center}
\end{mydef} 
where $diam(G)$ is the diameter of $G$ and $\delta$ is a user defined threshold. The generalization of peripheral vertices to pseudo-peripheral vertices provides a degree of freedom in order to find as many \textit{distant} nodes as we'll need later. 

\section{Community Detection}
\label{sec:comm}

\textit{Community detection} is a research branch in Network Science investigating the problem of finding similar vertices in a network according to some criteria. The name refers to the fact that similar vertices in a social network graph represent a \textit{community}. This problem is also referred to as \textit{Graph Clustering} in Computer Science literature. From data mining point of view, community detection is simply a \textit{clustering} task on the set of vertices of a graph. Clustering is one of the oldest topics in Machine Learning which aims to partition objects of a space into different groups such that the \textit{similarity} of objects within a cluster and \textit{dissimilarity} of objects between different clusters are maximized.

\subsection{Previous Works}
\label{sub::commlit}

However the problem of graph clustering is a relatively old problem in Mathematics \cite{fiedler1973algebraic,doi:10.1137/0130006} and Computer Science \cite{4766871,Urquhart1982173}, last decade has been a prominent period of time for this field with major contributions mostly from Physics community. The modern era of research in community detection started by the breakthrough idea of \textit{edge-betweenness} \cite{Girvan11062002} which brought forth an entire decade of remarkable research and significant results. Definition of edge-betweenness is based on a measure called node-betweenness proposed by Freeman in \cite{citeulike:1025135}. Edge-betweenness measures the importance of an edge in the process of flow of information across a network. 

Shortly after this work was published, Newman \& Girvan proposed \textit{modularity} function as a quality measure for good clusters \cite{PhysRevE.69.026113}. This idea quickly gained popularity, becoming the most widely-used performance measure for community detection in the course of the last decade. Moreover, modularity optimization has become the underlying idea for several different methods in the field \cite{CNM,blondel2008fuc,citeulike:1404492}.

Most of the clustering and community detection algorithms, including modularity-based algorithms, are based on maximizing the \textit{similarity} of objects \textit{within} a cluster while our proposed method concerns maximizing the \textit{dissimilarity} of objects \textit{between} different clusters. Fig.~\ref{fig:toynet} explains the philosophy behind this approach. In the context of social networks, the network shown in Fig.~\ref{fig:toynet} includes two individuals who have many mutual friends. Modularity-based methods end up with considering the whole network as a single community but our modularity-free (periphery-based) algorithm considers these two individuals as two different communities while mutual friends form a third community as they are equidistant from both peripheral individuals. Here, the latter makes more sense as it assigns two significantly popular guys to two different communities while modularity-based approaches fail to capture this \textit{distance}.

\begin{figure}[htb!]
  \centering
  \centerline{\includegraphics[width=8.5cm]{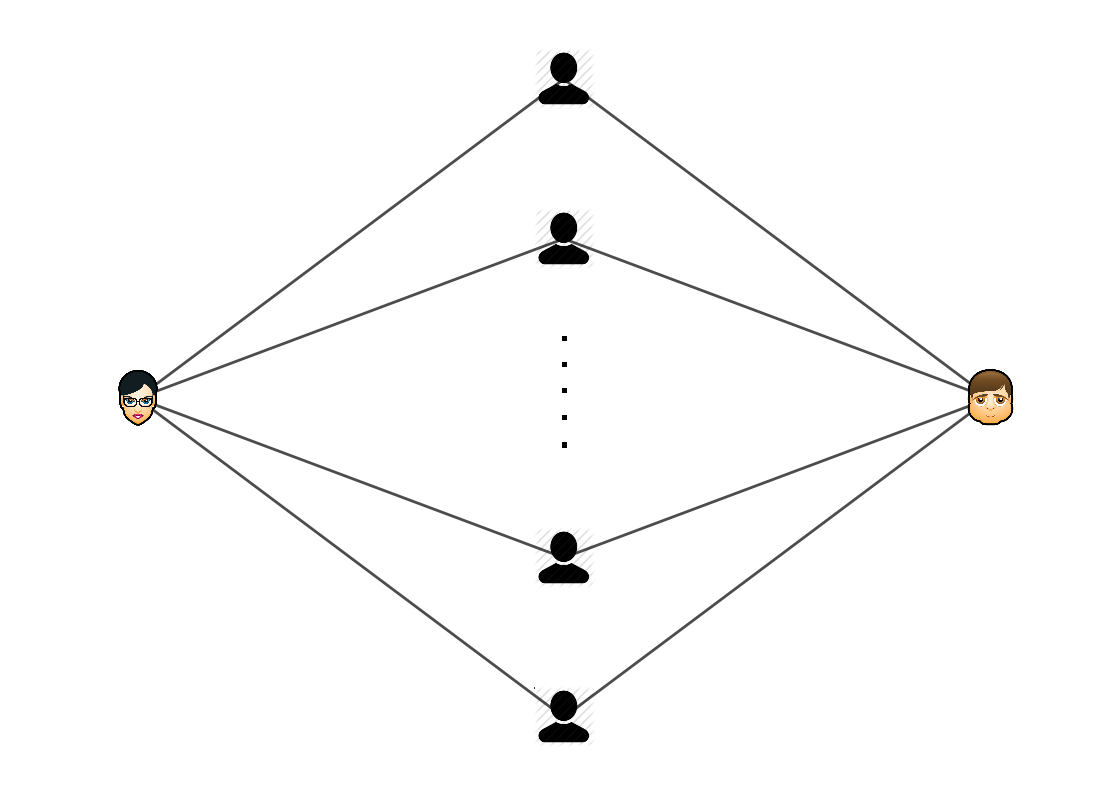}}
\caption{\label{fig:toynet}A toy social network in which two individuals share many mutual friends.}
\label{fig:toynet}
\end{figure}

Moreover, the underlying idea of community detection in complex networks needs to be revised as all traditional approaches focus on the concentration of relations as the main criterion (e.g. \textit{modularity}) while vertices can also be clustered due to their similar topological characteristics as well as their roles in the network. In Figure \ref{fig:toynet} all mutual vertices play the same role according to the flow of information from one peripheral vertex to another however, they are not adjacent.

\subsection{SGPD Volume Maximization Algorithm}
\label{sub:commalg}

The proposed algorithm is designed to select some landmark vertices in a graph (which will be the pseudo-peripheral vertices) as representatives of clusters and assign other vertices to these landmarks based on geodesic distances. The story-line of our proposed algorithm is simply as follows:
\begin{itemize}
 \item Finding $k$ pseudo-peripheral nodes
 \item Detecting ambiguous nodes
 \item Determining the community structure
\end{itemize}

In this section we describe each step in details.
\subsubsection{Finding Peripheral Vertices \& Their Communities}
\label{subsub:commper}
Here we explain an iterative algorithm to find pseudo-peripheral vertices in a graph based on SGPD (Spectral Graph Pseudo-peripheral and Pseudo-diameter) algorithm proposed in \cite{paulino}. According to the SGPD algorithm, the smallest and largest elements of the eigenvector associated to the second smallest eigenvalue of the Laplacian matrix of a graph (a.k.a. Fiedler vector) correspond to the endpoints of a diameter i.e. they are first two peripheral vertices $p_{1}$ and $p_{2}$ described in Observation \ref{obs:1}. Given these two peripheral vertices, we look for the next pseudo-peripheral node that maximizes the volume of virtual polygon shown in Fig.~\ref{fig:PerNodes} and also stands as equally far as possible from both first peripheral nodes. Algorithm~\ref{alg:sgpdvolmax} illustrates the procedure of \textit{SGPD Volume Maximization} method.

\begin{algorithm}[]
\SetAlgoLined
 \SetKwInOut{Input}{input}\SetKwInOut{Output}{output}
 \Input{Graph $G$, $k$ (number of pseudo-peripheral vertices)}
 \Output{$k$ pseudo-peripheral Vertices }
 PeriphVertices = SGPD(G);\\
 counter = 0;\\
 Distances = 0;\\
 \While{$counter \leq k$}{
 DistanceVectors $\leftarrow$ Single-source shortest path distance from each PeriphVertices to all other vertices;\\
 $\Theta$ $\leftarrow$ Compute the volume vector;\\
 Append(PeriphVertices, $\argmax_{v_{i}}\Theta$);\\
 counter $\leftarrow$ counter+1
 }
 \caption{\label{alg:sgpdvolmax}SGPD Volume Maximization}
\end{algorithm}

If we keep the calculated volume in a volume vector called $\Theta$, the $i^{th}$ element of such vector at $j^{th}$ iteration is defined as
\begin{equation}
\label{eq:volel}
\theta^{j}_{v_i}=\frac{{\displaystyle \prod_{k=1}^{j+1} d(v_{i},p_{k})}}{\sqrt{\displaystyle\sum_{k=1}^{j+1} d(v_{i},p_{k})^2}} \quad v_{i} \in V, j\geq1
\end{equation}
where $d(v,u)$ stands for the geodesic distance between nodes $v$ and $u$. All other pseudo-peripheral nodes after $p_{1}$ and $p_{2}$ are chosen as 
\begin{equation}
\label{eq:p3}
\qquad p_{j+2} = \argmax_{v_{i}}\{\theta^{j}_{v_i}\}
\end{equation}
respectively.

The volume vector $\Theta$ keeps an intuition of volume at each step and can be used for maximizing the volume in the next step. The reason we use the term \textit{volume}, as mentioned earlier, is that it is produced by multiplying \textit{lengths} thus intuitively reminds the concept of geometric volume. The denominator term in~\eqref{eq:volel} forces the next peripheral vertex to be as equidistant as possible from the first two while the enumerator maximizes the distance from them. After finding $k$ peripheral vertices we simply assign each vertex to its closest peripheral vertex. 
\subsubsection{Dealing With Ambiguous Vertices}
\label{subsub:AmbNodes}
\begin{figure*}[t!]
\centering
\subfloat[A Toy Barabási-Albert Network]{\includegraphics[width=0.32\textwidth]{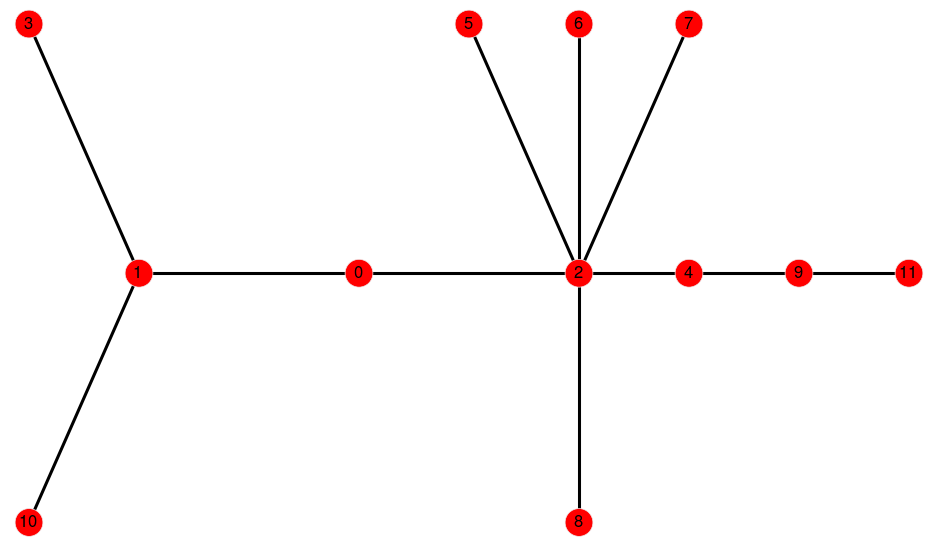}}{\ }
\subfloat[\label{fig:exp1b}Proposed Community Structure]{\includegraphics[width=0.32\textwidth]{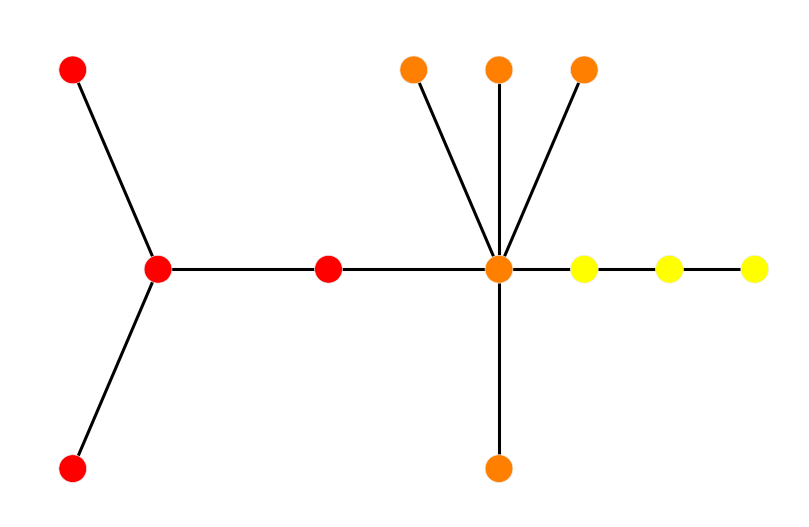}}{\ }
\subfloat[Blondel Community Structure]{\includegraphics[width=0.32\textwidth]{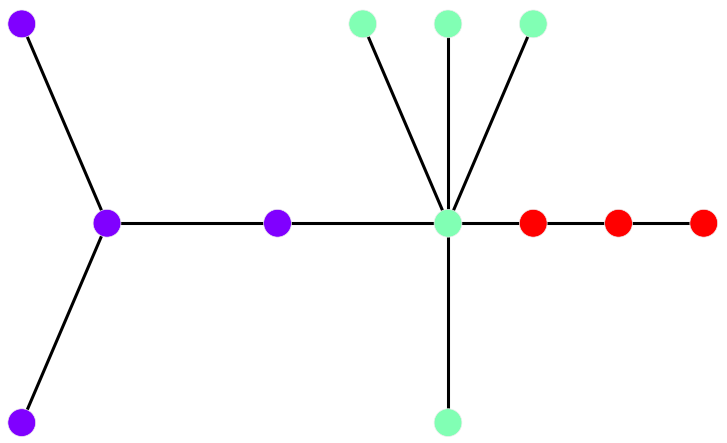}}{\ }
\caption{\label{fig:exp1} A demonstration of how the proposed algorithm copes with the \textit{ambiguous vertices}. \textit{(a)} shows a simple Barabási-Albert network \cite{Barabási15101999} which was partitioned into 2 clusters according to vertices $10$ and $11$ as peripheral vertices, attracting $red$ and $yellow$ vertices in \textit{(b)} respectively. Orange color in \textit{(b)} demonstrates ambiguous vertices which are equidistant from both community representatives thus form a third community. \textit{(c)} shows the result of Blondel algorithm \cite{blondel2008fuc} for community detection on this network.}
\end{figure*}

The main issue regarding this approach shows up when a vertex stands at the same distance from two or more different peripheral nodes (orange vertices in Fig.~\ref{fig:exp1b}). We call these nodes \textit{Ambiguous Vertices} from now on. Ambiguous vertices are of interest as they represent the most uncertainty of decision in our clustering approach and they construct a basis for soft clustering approach which is out of scope of this note. Moreover, these ambiguous vertices are the intermediate blocks between two or more endpoints thus directly affect the flow of information across the network. 

\begin{figure*}[t!]
\centering
\subfloat[$Q=0.72$]{\includegraphics[width=0.32\textwidth]{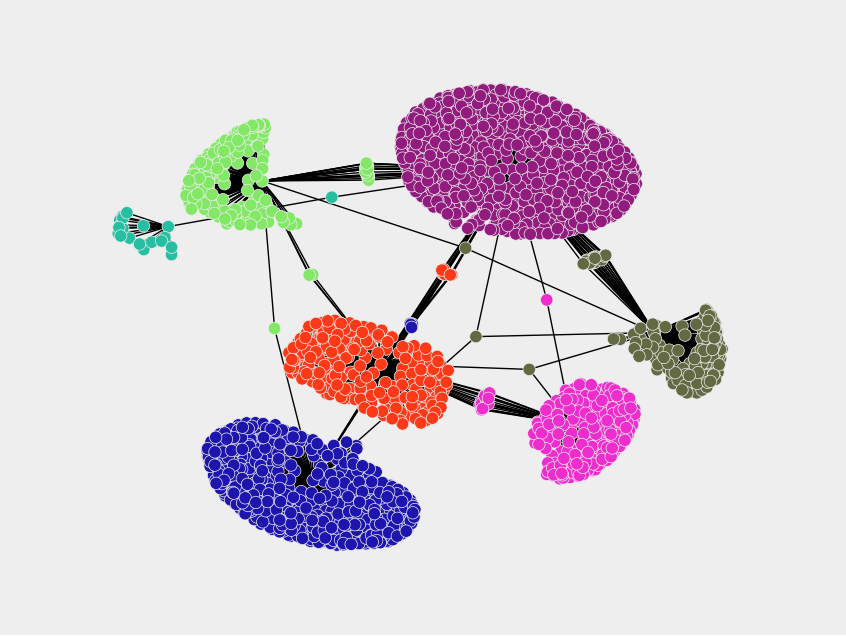}}{\ }
\subfloat[$Q=0.66$]{\includegraphics[width=0.32\textwidth]{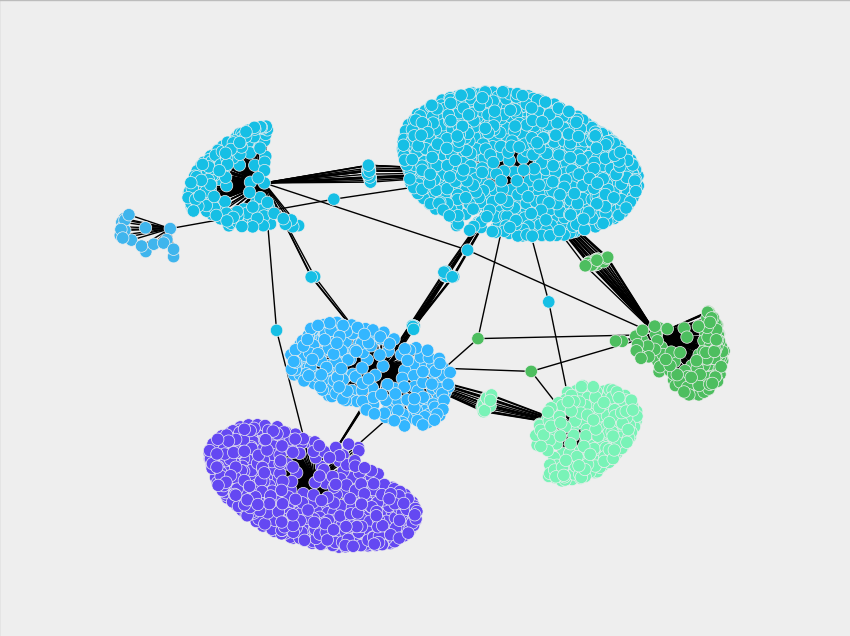}}{\ }
\subfloat[$Q=0.71$]{\includegraphics[width=0.32\textwidth]{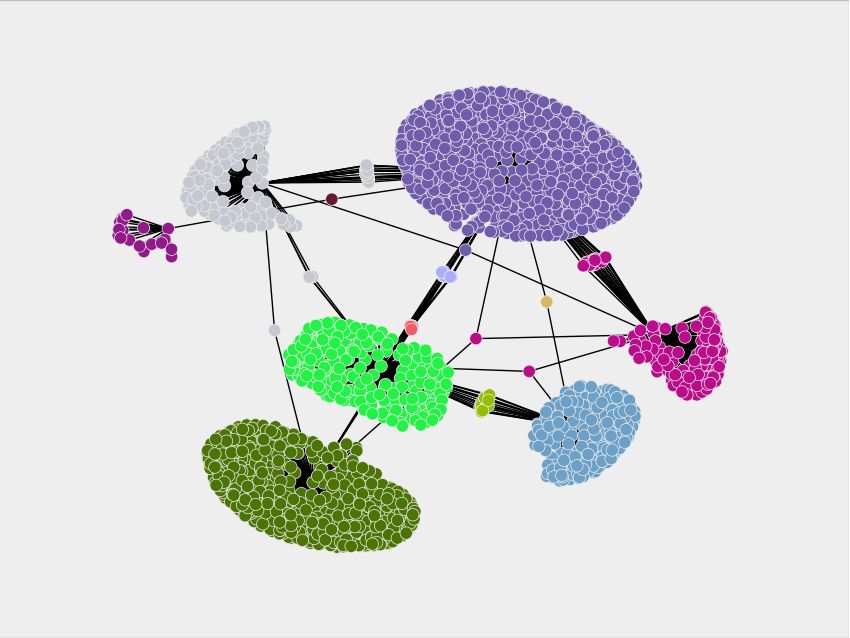}}{\ }
\caption{\label{fig:gplus} The result of \textit{(a)} Blondel algorithm, \textit{(b)} proposed algorithm before node disambiguation and \textit{(c)} the proposed algorithm after node disambiguation on a subgraph of \textit{Google+} social network \cite{Mcauley:2014:DSC:2582178.2556612} with 2400 nodes and 7 communities detected by Blondel algorithm. The modularity scores are reported as $Q$.}
\end{figure*}

As shown in Fig.~\ref{fig:exp1}, one can view ambiguous vertices as normal communities i.e. vertices that are at the same distance from two or more peripheral vertices are considered as a community themselves. In Fig.~\ref{fig:exp1}, we started with two communities and the algorithm automatically detected the third one based on ambiguous vertices. An illustration of SGPD Volume Maximization algorithm on a subgraph of \textit{Google+} social network is shown in Fig.~\ref{fig:gplus}.

\section{Experimental Results}
\label{sec:expcomm}
In this section an experimental framework will be set up to study different aspects of the proposed periphery-based community detection algorithm. All experiments have been done on a single machine running Windows 7 with \textit{i7-3632QM} core (2.2 GHz) and 6 GB memory. All algorithms were implemented in Python 2.7.6 using \textit{Networkx} library for graph computations.

\begin{figure*}[t!]
\centering
\subfloat[\label{fig:size1}8 communities]{\includegraphics[width=0.45\textwidth]{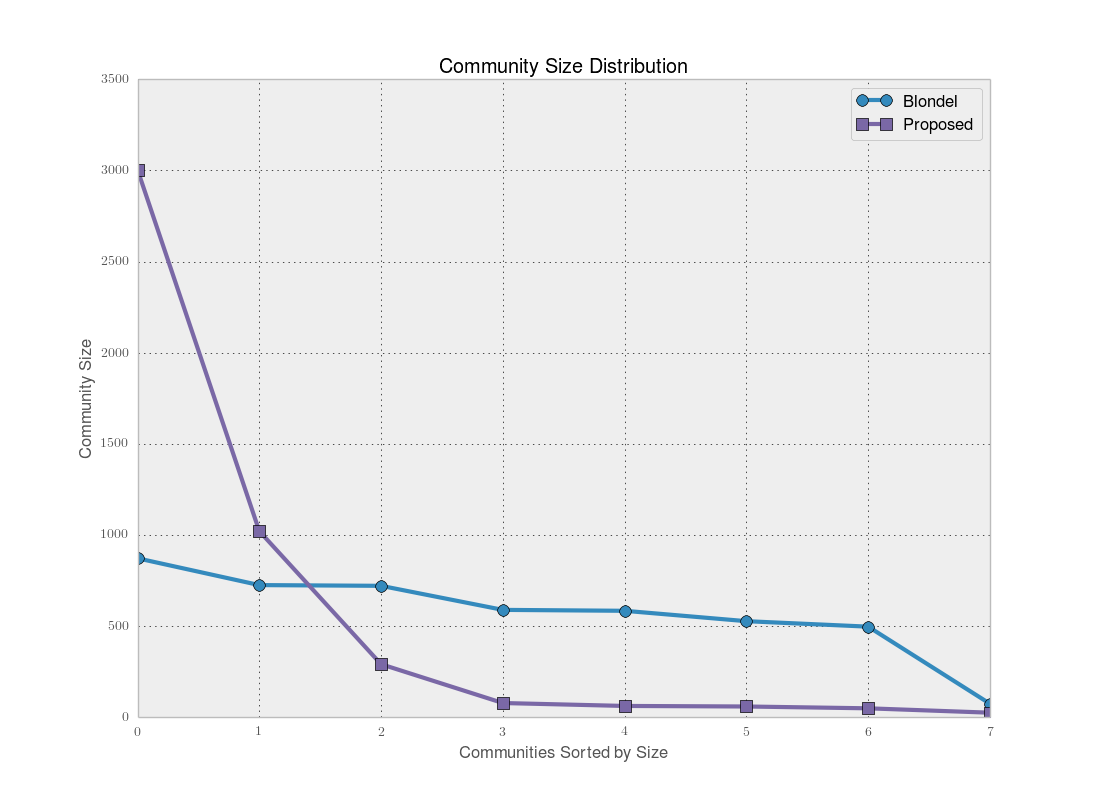}}{\ }
\subfloat[\label{fig:size2}146 communities]{\includegraphics[width=0.45\textwidth]{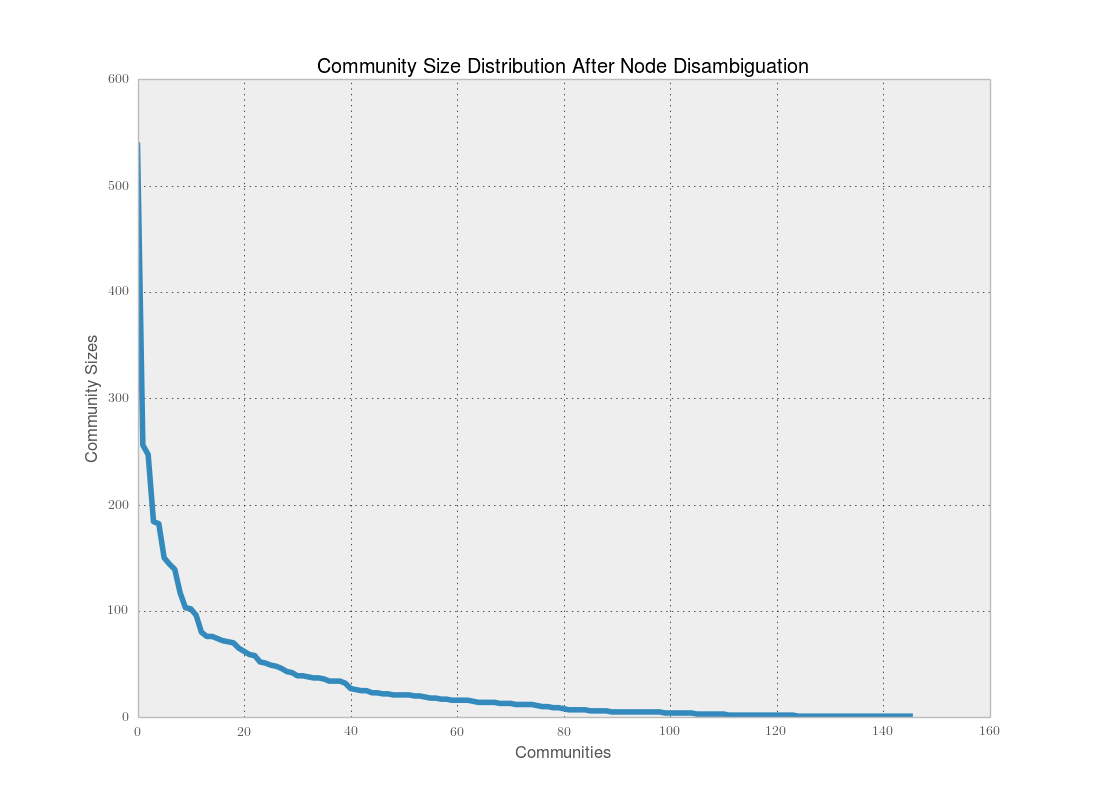}}{\ }
\caption{\label{fig:size} The distribution of community sizes after running \textit{(a)} proposed method without disambiguation and also Blondel algorithm and \textit{(b)} our proposed method after disambiguation of vertices.}
\end{figure*}

The real-world network we consider here is a subgraph of Wikipedia graph collected through Wikispeedia online game \cite{wikispeedia1}, \cite{wikispeedia2}. This network contains 4589 vertices and 106,644 unweighted edges. The original path traversing graph is directed but we only consider the connections regardless of the direction to make the graph suitable for our algorithm. Figure \ref{fig:size} shows the distribution of size of communities in Blondel algorithm, our proposed method before disambiguation and after disambiguation. For the sake of comparison, the Blondel algorithm was used in our experiments as a well-established modularity based method. 

Figure \ref{fig:size1} implies the difference between community sizes in modularity-based and periphery-based approaches to community detection. Blondel algorithm has captured almost balanced communities while the size of communities in our method obeys the structure of the network according to its power-law nature. It depicts the fact that periphery-based method does not put two vertices in a same cluster according to the concentration of edges around them, rather the vertices are placed according to their distances to the periphery of the graph i.e. the closer a community gets to the boundary of the network, the smaller the size of community becomes. This is expected as the degree of nodes decreases by getting far from the central vertices so the probability of two vertices being in the same cluster decreases. 

\begin{figure}[htb!]
  \centering
  \centering{\includegraphics[width=0.45\textwidth]{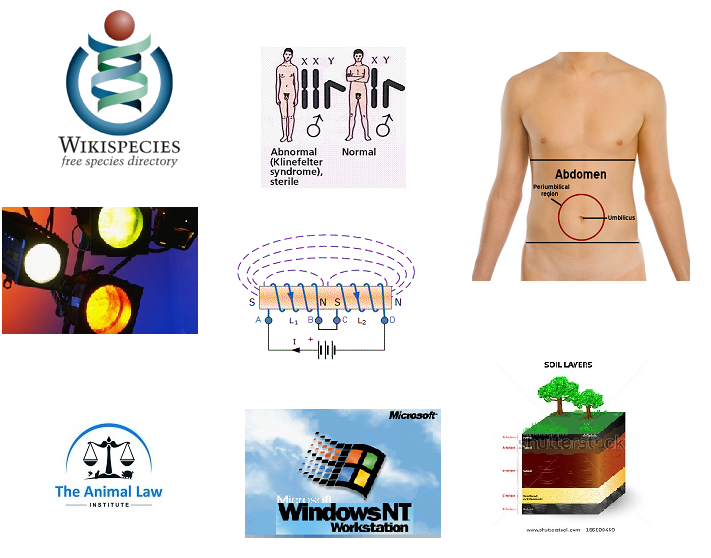}}
\caption{\label{fig:wiki}\textit{Wikispecies}, \textit{Klinefelter syndrome}, \textit{human abdomen}, \textit{lighting designer}, \textit{inductance}, \textit{animal law}, \textit{Windows NT architecture} and \textit{soil profile} are the eight peripheral vertices of our Wikipedia subgraph.}
\label{fig:wiki}
\end{figure}

The peripheral vertices (Wikipedia pages) found by our algorithm are shown in Figure \ref{fig:wiki}. To understand the concept of periphery-based community detection better, we look at the topological position of \textit{"Diego Maradona"} in this subgraph of wikipedia as a case study. Blondel algorithm puts \textit{Maradona} beside 724 other vertices including many conceptually related Wikipedia pages as well as many irrelevant pages. Modularity maximization of the Blondel algorithm concerns the density of relevant links but here it does not make sense as \textit{Maradona} is not conceptually related to \textit{Super Mario 64} or \textit{Rail Transport}! This simple example shows the idea behind using peripheral distances as the criterion for clustering vertices based on the \textit{topological} mutual characteristic. Our proposed method assigns \textit{Diego Maradona} to a cluster along with 70 other vertices which are all conceptually irrelevant but topologically stand at the same geodesic distance from peripheral vertices \textit{human abdomen}, \textit{Wikispecies}, \textit{architecture of Windows NT} and \textit{lighting designer}. This result is reliable for understanding the role of vertices which block the flow of information between boundaries of network. The last but not least is the fact that as the degree distribution becomes more power-law, our result gets closer to those of modularity-based algorithms.

\section{Conclusion}
\label{sec:conc}

We briefly proposed the idea of community detection using peripheral vertices instead of center-based approaches. This approach to community structure prevents getting stuck in modularity maximization constraints (e.g. resolution of the clustering) and also levers a basis for understanding the topological roles that different vertices and communities play in the network. These topological properties can be a reliable characteristic for axiomizing the graph clustering problem as it does not vary with the attributes of nodes. This approach also works pretty well on power-law graphs which is the main structure of real-world complex networks. 

As a direction for further investigation we are going to reduce the computational complexity of finding peripheral vertices via sampling vertices according to the distribution of shortest path lengths. Optimizing an objective function also seems necessary to provide a basis for theoretical analysis of such an algorithm and also to find the optimal number of clusters as the proposed method needs it a priory. The proposed algorithm also opens a new window to the problem of graph embedding as it captures the topological structure of geodesic distances in the network.

\bibliographystyle{IEEEtran}
\bibliography{Bib}

\end{document}